%% file: main.tex
\documentclass{article}

\usepackage{amsmath}
\usepackage[american]{babel}
\usepackage{graphicx}
\usepackage[utf8x]{inputenc}
\usepackage{dsfont}
\usepackage{subfigure}
\usepackage{todonotes}
\usepackage{xspace}
\usepackage{url}
\usepackage{wrapfig}
\usepackage{placeins}

\graphicspath{{./fig/}}

\newcommand{\bigO}{\operatorname{O}}

\newcommand{\R}{\mathds{R}}
\newcommand{\V}{\mathcal{V}}

\newcommand{\email}[1]{{\tt #1}}
\newcommand{\ignore}[1]{}

\title{Efficient Computation of Visibility Polygons}

\author{
  Francisc Bungiu\thanks{Institute of Information Security at ETH Z\"urich, Switzerland. \email{fbungiu@gmail.com}} \and 
  Michael Hemmer\thanks{Department of Computer Science, TU Braunschweig, Germany. \email{mhsaar@gmail.com}, \email{kroeller@ibr.cs.tu-bs.de}}  \and 
  John Hershberger\thanks{Mentor Graphics Corp., Wilsonville, OR, \mbox{USA}. 
        \email{john\_hershberger@mentor.com}} \and 
	Kan Huang\thanks{Department of Applied Mathematics \& Statistics, Stony Brook University, New York, USA. \email{huangkandiy@gmail.com}} \and 
  Alexander Kr\"oller$^\dagger$}

\begin{document}

\maketitle

\input{abstract}
\input{introduction}
\input{related_work}
\input{algorithms}

\input{experiments}

\paragraph{Acknowledgements.}
\FloatBarrier
This work was supported by {\em Google Summer of Code} and the Deu\-tsche For\-schungs\-ge\-mein\-schaft~(DFG), contract KR~3133/1-1 (Kunst!).

{\small
\bibliographystyle{abbrv}
\bibliography{bibliography}
}

\end{document}

%% file: abstract.tex
\begin{abstract}
  Determining visibility in planar polygons and arrangements is
  an important subroutine for many algorithms in computational
  geometry. In this paper, we report on new implementations, and
  corresponding experimental evaluations, for two established and one
  novel algorithm for computing visibility polygons. These algorithms
  will be released to the public shortly, as a new package for the
  Computational Geometry Algorithms Library (CGAL). 
\end{abstract}


%% file: introduction.tex
\section{Introduction}
\label{sec:introduction}

\begin{wrapfigure}[10]{r}{0.3\textwidth}
  \centering
  \includegraphics[width=.3\textwidth]{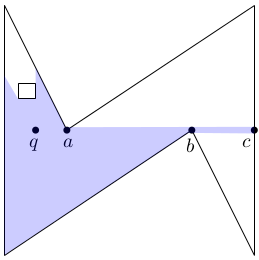}
  \label{fig:antennae}
\end{wrapfigure}

Visibility is a basic concept in computational geometry. For a
polygon $P\subset\R^2$, we say that a point $p\in P$ is {\em visible}
from $q\in P$ if the line segment $\overline{pq}\subseteq P$.
The points that are visible from $q$ form the visibility region
$\V(q)$. Usually (if no degeneracies occur) $\V(q)$ is a polygon;
hence it is often called the {\em visibility polygon}. If there
are multiple vertices collinear with $q$, then
$\V(q)$ may take the form of a polygon with attached one-dimensional
antennae (see Figure~\ref{fig:antennae}).

Computing $\V(q)$, given a query point $q$, is a
well-known problem with a number of established algorithms; see the
book by Ghosh~\cite{ghosh2007visibility}. Besides being of interest on
its own, it also frequently appears as a subroutine in algorithms for
other problems, most prominently in the context of the Art Gallery
Problem~\cite{o-agta-87}. In experimental work on this
problem~\cite{kbfs-esbgagp-12} we have identified
visibility computations as having a substantial impact on
overall computation times, even though it is a low-order
polynomial-time subroutine in an algorithm solving an NP-hard
problem. Therefore it is of enormous interest to have efficient
implementations of visibility polygon algorithms available.

CGAL, the Computational Geometry Algorithms Library~\cite{cgal},
contains a large number of algorithms and data structures, but
unfortunately not for the computation of visibility polygons. We present
implementations for three algorithms, which form the basis for an
upcoming new CGAL package for visibility. Two of these are efficient
implementations for standard $\bigO(n)$-~and $\bigO(n\log n)$-time
algorithms from the literature. In addition, we describe a novel
algorithm with preprocessing, called {\em triangular expansion}, 
based on an intriguingly simple idea using a triangulation of 
the underlying polygon. Even though this
algorithm has a worst-case query runtime of $\bigO(n^2)$, it is
extremely fast in practice. In experiments we have found it to
outperform the other implementations by about two orders of magnitude
on large and practically relevant inputs. 
The implementations follow the exact computation 
paradigm and handle all degenerate cases, whether or not to include
antennae in the output is configurable. 
Such a visibility polygon $\V(q)$ with antenna $bc$ is depicted on 
in the figure to the right. 

Our contribution is therefore threefold:  
We present a new algorithm for computing visibility
polygons. We report on an experimental
evaluation, demonstrating the efficiency of the implementations,
and revealing superior speed of the novel algorithm.
The publication of the code is scheduled for CGAL release~4.5 (October 2014).  



%% file: related_work.tex
\section{Related Work}
For a detailed coverage of visibility algorithms in two dimensions, 
see the book~\cite{ghosh2007visibility}.
We denote by $n$ the number of vertices 
of the input polygon $P$, by $h$ the number of its holes, and by 
$k$ the complexity of the visibility region $\V(q)$ of the query point 
$q$. 

The problem of computing $\V(q)$ was first addressed for simple polygons
in~\cite{davis1979computational}. The first correct $O(n)$ time 
algorithm was given by Joe and Simpson~\cite{joe1987corrections}.
For a polygon with holes, $O(n\log n)$ time algorithms were proposed by Suri 
et al.~\cite{suri1986worst} and Asano et al.~\cite{asano1986visibility}. 
An optimal $O(n+ h \log h)$ algorithm was later given by 
Heffernan and Mitchell~\cite{heffernan1995optimal}.

Ghodsi et al.~\cite{guibas1997robot} reduced the query time for simple
polygons to $O(\log n + k)$, at the expense of preprocessing requiring
$O(n^3)$ time and space. In~\cite{bose2002efficient} Bose et
al.~showed that this time can also be achieved for points outside of
$P$
with an $O(n^3\log n)$ preprocessing time. 
Aronov et al.~\cite{aronov2002visibility} reduced the preprocessing time 
and space to $O(n^2\log n)$ and $O(n^2)$ respectively with 
the query time being increased to $O(\log^2 n + k)$.   

Allowing preprocessing for polygons with holes,
Asano et al.~\cite{asano1986visibility}
presented an algorithm with $O(n)$ query time that 
requires $O(n^2)$ preprocessing time and space.
Vegter~\cite{vegter1990visibility} 
gave an output sensitive algorithm whose query time is $O(k\log (n/k))$,
where preprocessing takes $O(n^2 \log n)$ time and $O(n^2)$ space. 
Zarei and Ghodsi~\cite{zarei2005efficient} gave another output 
sensitive algorithm whose query time is $O(1+\min(h,k)\log n +k)$ 
with $O(n^3\log n)$ preprocessing time and $O(n^3)$ space. 
Recent work by Inkulu and Kapoor~\cite{inkulu2009visibility} extends
the approaches of~\cite{aronov2002visibility}
and~\cite{zarei2005efficient} by incorporating a trade-off between 
preprocessing and query time.

Surprisingly, it seems that there are no exact and complete 
implementations of these algorithms at all.  
The only available software is GEOMPACK~\cite{joe1993geompack} 
but it does not use exact arithmetic and only implements the algorithm 
for simple polygons of Joe and Simpson~\cite{joe1987corrections}.


%% file: algorithms.tex
\section{Implemented Algorithms}
\label{sec:algos}

The following algorithms are implemented in {\tt C++}, and are part of an upcoming 
CGAL package for visibility. 
\input{simple}
\input{rotational_sweep}
\input{triangular_expansion}

%% file: simple.tex
\subsection{Algorithm of Joe and Simpson~\cite{joe1987corrections}}
\label{ssec:simple_algo}

The visibility polygon algorithm by Joe and Simpson~\cite{joe1987corrections}
runs in $O(n)$ time and space.  It performs a sequential scan of the boundary
of $P$, a simple polygon (without holes), and uses a stack $s$ of boundary points 
$s_0, s_1,\ldots,s_t$ such that at any time, the stack represents the visibility
polygon with respect to the already scanned part of $P$'s boundary.

When the current edge $v_iv_{i+1}$ is in process, three possible 
operations can be performed: new boundary points are added to the stack,
obsolete boundary points are deleted from the stack, or subsequent
edges on the polygon's boundary are scanned until a certain condition is met.
The later is performed when the edge enters a so-called ``hidden-window,''
where the points are not visible from the viewpoint $q$.

The implementation uses states to process the boundary of the
polygon $P$ and manipulate the stack $s$ such that, in the end, it contains
the vertices defining the visibility polygon's boundary. The main state
handles the case when the current edge $v_iv_{i+1}$ takes a left turn
and $v_{i+1}$ is pushed on the stack. The other states are entered
when $v_iv_{i+1}$ turns behind a previous edge. Then the algorithm
scans the subsequent edges until the boundary reappears from behind
the last visible edge. In order to deal with cases in which the polygon
winds more than $360^\circ$, a winding counter is used during this
edge processing.



%% file: rotational_sweep.tex
\subsection{Algorithm of Asano~\cite{asano1985efficient}}
\label{ssec:rotational_sweep}

The algorithm by Asano~\cite{asano1985efficient} runs in $O(n\log n)$
time and $O(n)$ space.
It follows the classical plane sweep paradigm with 
the difference that the event line $L$ is a ray  that originates from 
and rotates around the query point $q$. Thus, initially all vertices 
of the input polygon (event points) are sorted according to their 
polar angles with respect to the query point $q$. 
Segments that intersect the ray $L$ are stored in a balanced 
binary tree $T$ based on their order of intersections with $L$. 
As the sweep proceeds, $T$ is updated and a new vertex of $V(q)$ 
is generated each time the smallest element (segment closest to $q$) 
in $T$ changes.

The algorithm can be implemented using basic algorithms and 
data structures such as {\tt std::sort} (for ordering event points)
and {\tt std::map} (ordering segments on the ray). 
However, it is crucial that the underlying comparison operations 
are efficient. We omit details of the angular comparison for vertices 
and sketch the ordering along $L$. 

\ignore{To compare the angular order of two vertices we 
detect the quadrant of each vertex with respect to $q$ 
(comparing coordinates of $x$ and $y$), which often already 
determines the order.
In case both vertices are in the same quadrant this is 
followed by one call to the standard orientation predicate. }

\begin{wrapfigure}[12]{r}{0.4\textwidth}
\vspace{-0.33cm}
\includegraphics[width=0.4\textwidth]{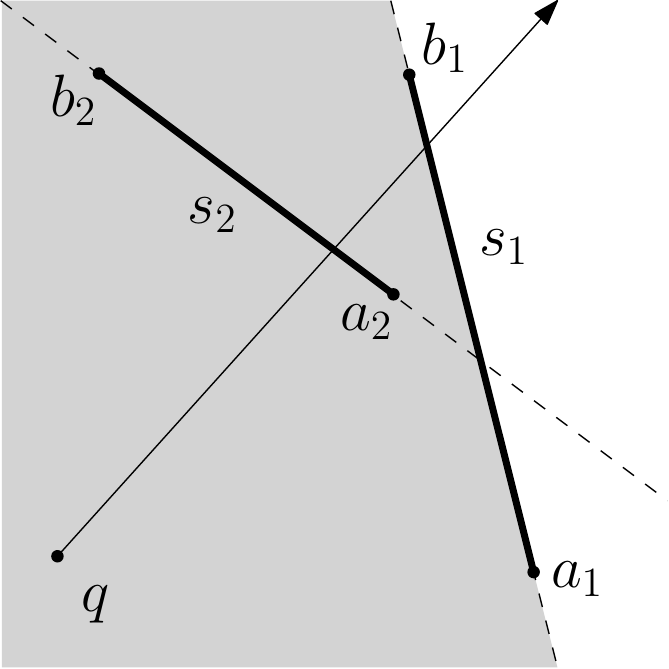}
\end{wrapfigure}
For an efficient comparison of two segments $s_1$ and $s_2$ 
along the ray $L$, it is important to avoid explicit 
(and expensive) construction of intersection points with $L$. 
In fact, $L$ can stay completely abstract. 
Assuming that $s_1$ and $s_2$ have no common endpoint, the order 
can be determined using at most five orientation predicates. 
Consider the figure to the right: 
the segment $s_1$ is further away since $q$, $a_2$ and $b_2$ are 
on the same side of $s_1$ (three calls). In case this is not conclusive, 
two additional calls for $q$ and $a_1$ with respect to $s_2$ 
are sufficient.

\ignore{
The Algorithm by T.Asano~\cite{asano1985efficient} is an $O(n\log n)$ time 
algorithm based on angular plane sweep. All vertices of the polygonal
 environment are sorted according to their polar angles at query point 
$q$. Then rotate a ray $L$ around $q$. The segments intersected by the ray
are stored in a balanced binary tree $T$ based on their orders of 
intersections with $L$. As the sweep proceeds, $T$ will be updated and 
a new vertex of $V(q)$ is detected as soon as the smallest 
element (segment closest to q) in $T$ changes. 
Therefore after a round of plane sweep, $V(q)$ is obtained. 
The algorithm requires $O(n)$ space.
}

%% file: triangular_expansion.tex
\subsection{New Algorithm: Triangular Expansion}
\label{ssec:triangular_expansion}

We introduce a new algorithm that, to the best of our knowledge, has not 
been discussed in the literature. 
Its  worst case complexity is $O(n^2)$ but it is very 
efficient in practice as it is, in some sense, output sensitive. 

In an initial preprocessing phase the polygon is triangulated. 
For polygons with holes this is possible in $O(n\log n)$ 
time~\cite{dutchbook} and in $O(n)$ time for simple 
polygons~\cite{chazelle-sptri}. However, for the implementation
we used CGAL's constrained Delaunay triangulation which requires
$O(n^2)$ time in the worst case but behaves well in practice. 

\begin{figure}[h]
  \centering
  \includegraphics[width=0.7\textwidth]{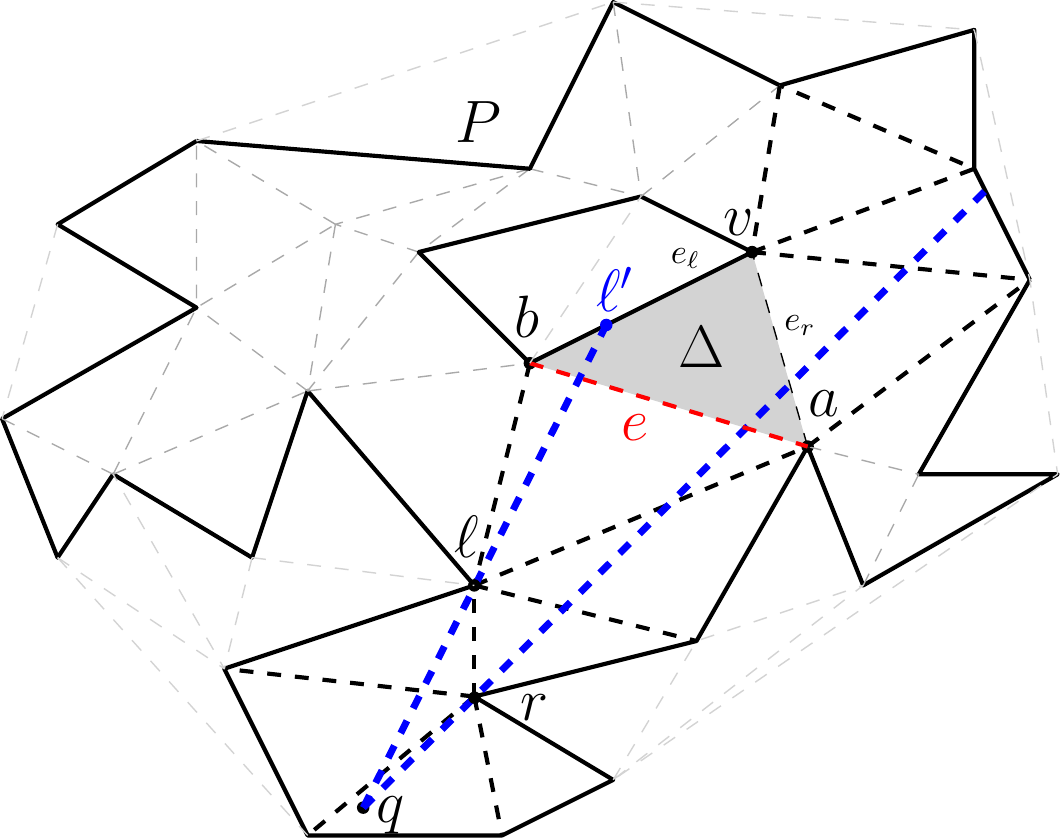}
  \caption{Triangular expansion algorithm --- recursion entering triangle $\Delta$ through edge $e$.}
  \label{fig:trianglular_expansion}
\end{figure}

Given a query point $q$, we locate the triangle 
containing $q$ by a simple walk.
Obviously $q$ sees the entire triangle and we can report
all edges of this triangle that belong to $\partial P$.
For every other edge we start a recursive procedure that expands
the view of $q$ through that edge into the next triangle.
Initially the view is restricted by the two endpoints of the edge. However, 
while the recursion proceeds the view can be further restricted. This is depicted in Figure~\ref{fig:trianglular_expansion}:
The recursion enters triangle $\Delta$ through edge $e$ with endpoints
$a$ and $b$. 
The view of $q$ is restricted by the two 
reflex vertices $l$ and $r$, where $a \leq r < \ell \leq b$ with respect to 
angular order around $q$. The only new vertex is $v$ and its position with 
respect to $\ell$ and $r$ is computed with two orientation tests. 
In the example of Figure~\ref{fig:trianglular_expansion} the vertex $v$ 
is between $\ell$ and $r$, thus we have to consider both edges $e_\ell$ and 
$e_r$: $e_\ell$ is a boundary edge and we can report edge $\overline{\ell\ell'}$
and $\overline{\ell'v}$ as part of $\partial{\V(q)}$; $e_r$ is not a boundary 
edge, which implies that the recursion continues with $v$ being the 
vertex that now restricts the left side of the view.

The recursion may split into two calls if $e_\ell$ and $e_r$ 
are both not part of the boundary. 
As there are $n$ vertices, this can happen 
$O(n)$ times; each call may reach $O(n)$ triangles, which suggests a
worst-case runtime of $O(n^2)$. However, a true split into two visibility 
cones that may reach the same triangle independently can happen only at 
a hole of $P$. Thus, at worst the runtime is $O(nh)$, where $h$ is the 
number of holes. This implies that the runtime is linear for simple polygons.

\begin{figure}[h]
  \centering
  \def\svgwidth{0.7\textwidth}
  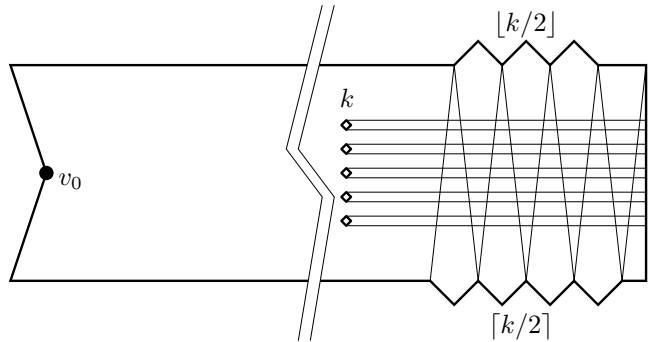
  \caption{$\Theta(n^2)$ worst case for triangular expansion.}
  \label{fig:triangwc}
\end{figure}

Figure~\ref{fig:triangwc} sketches the worst case scenario, which we also used in 
our experiments (Section~\ref{sec:experiments}). The $k =
\Theta(n)$ holes in the 
 middle of the long room split the view of $v_0$ into $k+1$ cones, each of 
which passes through $k = \Theta(n)$ Delaunay edges. 
Thus, in this scenario the algorithm has complexity $\Theta(n^2)$. 

However, ignoring the preprocessing, the algorithm often runs in
sublinear time, since it processes only those triangles that are actually seen. 
This can be a very small subset of the actual polygon. Unfortunately, this
is not strictly output sensitive, since a visibility cone may 
traverse a triangle even though the triangle does not contribute to
the boundary 
of $\V(q)$. 

Since the triangulation has linear size and since 
at any time there are at most $O(n)$ recursive calls on the stack,
the algorithm requires $O(n)$ space. 


%% file: fig/triang-worst-case.pdf_tex
\begingroup%
  \makeatletter%
  \providecommand\color[2][]{%
    \errmessage{(Inkscape) Color is used for the text in Inkscape, but the package 'color.sty' is not loaded}%
    \renewcommand\color[2][]{}%
  }%
  \providecommand\transparent[1]{%
    \errmessage{(Inkscape) Transparency is used (non-zero) for the text in Inkscape, but the package 'transparent.sty' is not loaded}%
    \renewcommand\transparent[1]{}%
  }%
  \providecommand\rotatebox[2]{#2}%
  \ifx\svgwidth\undefined%
    \setlength{\unitlength}{212.95bp}%
    \ifx\svgscale\undefined%
      \relax%
    \else%
      \setlength{\unitlength}{\unitlength * \real{\svgscale}}%
    \fi%
  \else%
    \setlength{\unitlength}{\svgwidth}%
  \fi%
  \global\let\svgwidth\undefined%
  \global\let\svgscale\undefined%
  \makeatother%
  \begin{picture}(1,0.52629725)%
    \put(0,0){\includegraphics[width=\unitlength]{triang-worst-case.pdf}}%
    \put(0.07771777,0.24435433){\color[rgb]{0,0,0}\makebox(0,0)[lb]{\smash{$v_0$}}}%
    \put(0.8102841,0.48585968){\color[rgb]{0,0,0}\makebox(0,0)[b]{\smash{$\lfloor k/2\rfloor$}}}%
    \put(0.80223394,0.01224062){\color[rgb]{0,0,0}\makebox(0,0)[b]{\smash{$\lceil k/2\rceil$}}}%
    \put(0.52986954,0.36644872){\color[rgb]{0,0,0}\makebox(0,0)[b]{\smash{$k$}}}%
  \end{picture}%
\endgroup%

%% file: experiments.tex
\section{Experiments}
\label{sec:experiments}

\begin{figure}[h] 
\centering
\includegraphics[width=.7\textwidth]{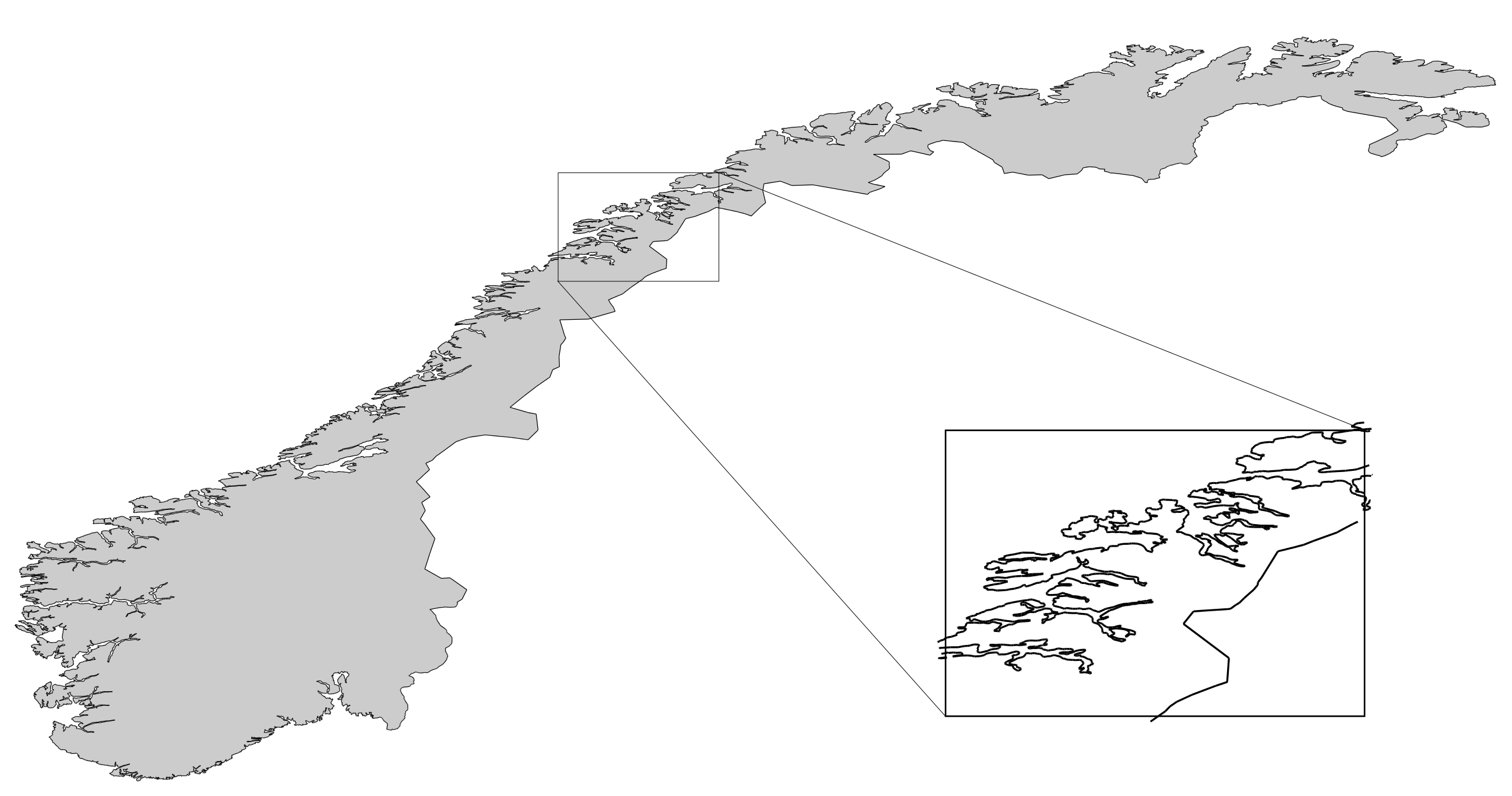}
\begin{tabular}{crrrr}
Alg.  & $T_{\mbox{\scriptsize \em PrePro}}$ & $T_{\mbox{\scriptsize \em Queries}}$  & $T_{\mbox{\scriptsize \em Total}}$ & {\em Avg} \\
S &  ---        &   117.43 s     &  117.43 s & 5.60 ms\\  
R &  ---        &  1193.29 s     & 1193.29 s & 56.87 ms\\  
T &  0.21 s     &     3.66 s     &    3.88 s & 0.18 ms\\
\end{tabular}
\caption{Norway example.}
\label{fig:norway}
\end{figure}

\begin{figure}[h]
\centering
\includegraphics[width=.7\textwidth]{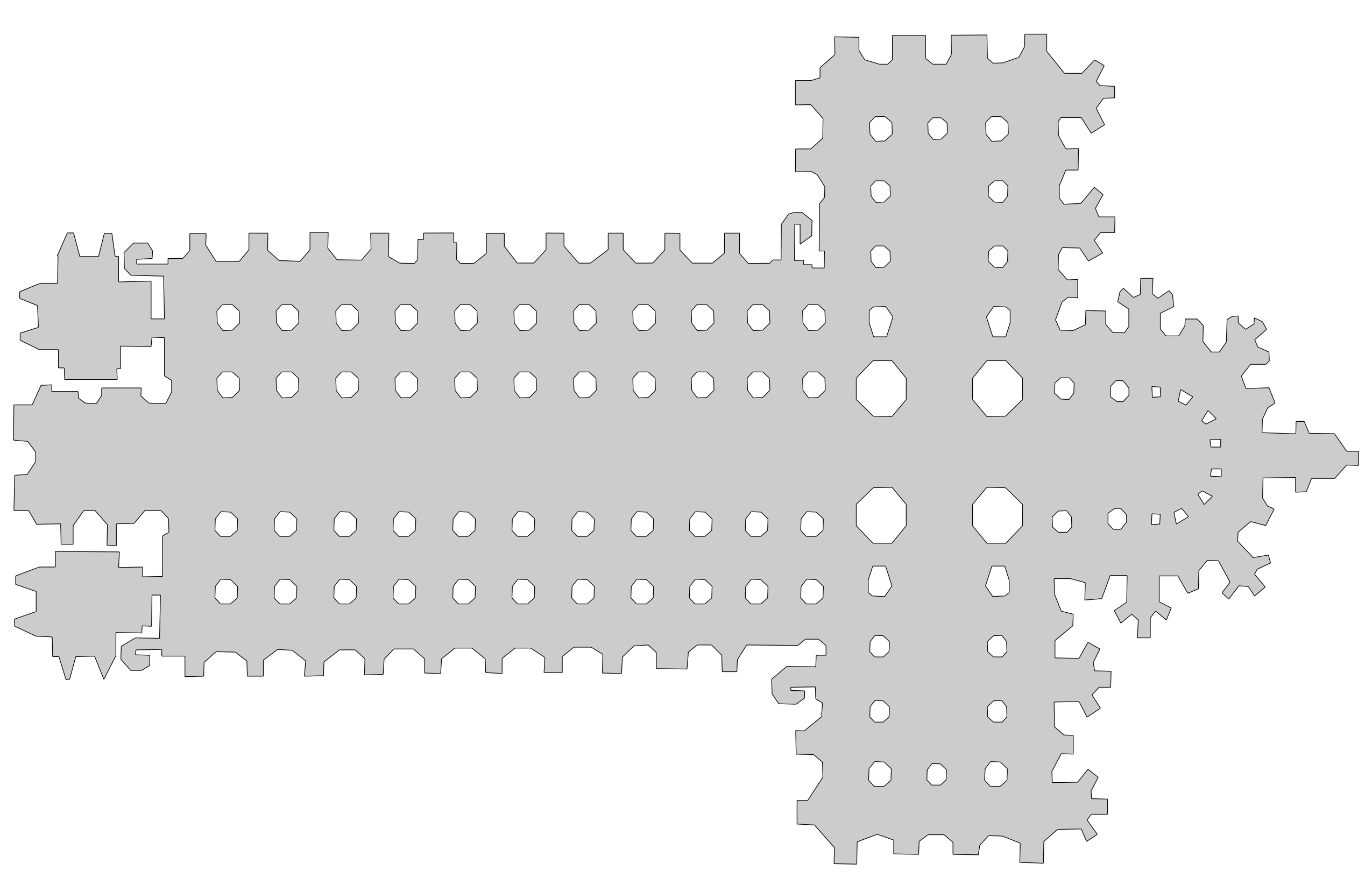}
\begin{tabular}{crrrr}
Alg.  & $T_{\mbox{\scriptsize \em PrePro}}$ & $T_{\mbox{\scriptsize \em Queries}}$  & $T_{\mbox{\scriptsize \em Total}}$ & {\em Avg} \\
R &  ---           & 1.35 s      &1.35 s &  1.112 ms\\  
T &  0.004 s       & 0.04 s      &0.04 s &  0.004 ms\\
\end{tabular}
\caption{Cathedral example.}
\label{fig:cathedral}
\end{figure}

The experiments were run on an Intel(R) Core(TM) 
i7-3740QM CPU at 2.70GHz with 6 MB cache and 
16 GB main memory running a 64-bit Linux  3.2.0 kernel. 
All algorithms are developed against the latest release (i.e.,~4.3) 
of CGAL. None of the algorithms uses parallelization.  
We tested three different scenarios: 
Norway (Figure~\ref{fig:norway}) a simple polygon 
with $20981$ vertices, 
cathedral (Figure~\ref{fig:cathedral}) a general polygon with 1209 vertices,
and the already mentioned worst case scenario (Figure~\ref{fig:triangwc}) 
for the triangular expansion algorithm. 

In tables $T_{Queries}$ indicates the total time to compute the
visibility polygons for all vertices, while $T_{total}$ also 
includes the time for preprocessing. $Avg$ indicates the average 
time required to compute the visibility area for one vertex 
including preprocessing. 
Algorithms are indicated as follows: 
(S) the algorithm of Joe and Simpson~\cite{joe1987corrections} for simple polygons;
(R) the algorithm of Asano~\cite{asano1985efficient} performing the rotational sweep around the query point;
(T) our new triangular expansion algorithm.

Since runtimes did not differ significantly 
we do not report on similar benchmarks with query points on edges 
and in the interior polygon.




\begin{figure}[h]
  \centering
  \includegraphics[width=0.5\textwidth]{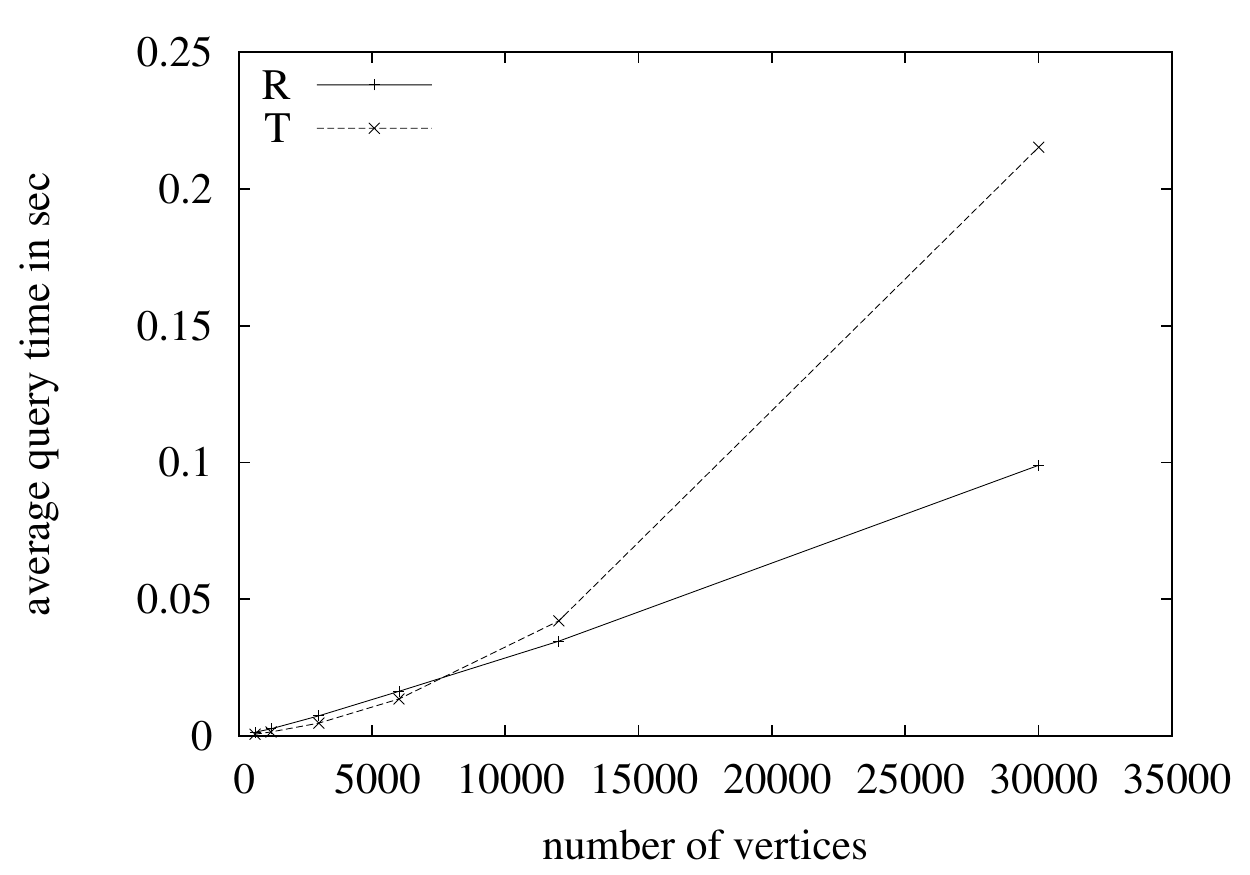}
  \caption{Worst case scenario. }
  \label{fig:exp_worst_case}
\end{figure}

For the real world scenarios, cathedral and Norway, we can observe that the average runtime 
of the triangular expansion algorithm is two orders of magnitude
faster than the rotational sweep algorithm.
It is more than one order of magnitude faster than the special
algorithm for simple polygons on the Norway data set.
However, for the worst case scenario,
Figure~\ref{fig:exp_worst_case} shows that eventually the sweep
algorithm becomes faster 
with increasing input complexity.


